\documentclass[dvips]{article}
\usepackage{graphicx}
\usepackage{amssymb}
\usepackage{graphicx}
\usepackage{dcolumn}
\usepackage{amsmath}
\usepackage{lscape}
\usepackage{longtable}
\usepackage{subfigure}
\usepackage{color}

\begin{document}
\newcommand{\bd}{\begin{document}}
\newcommand{\ed}{\end{document}}
\newcommand{\bc}{\begin{center}}
\newcommand{\ec}{\end{center}}
\newcommand{\bfr}{\begin{flushright}}
\newcommand{\efr}{\end{flushright}}
\newcommand{\lt}{\left}
\newcommand{\rt}{\right}
\newcommand{\vs}{\vspace}
\newcommand{\hs}{\hspace}
\newcommand{\beq}{\begin{equation}}
\newcommand{\eeq}{\end{equation}}
\newcommand{\lb}{\linebreak}
\newcommand{\pb}{\pagebreak}
\newcommand{\mb}{\makebox}
\newcommand{\fb}{\framebox}
\newcommand{\mc}{\multicolumn}
\newcommand{\ben}{\begin{enumerate}}
\newcommand{\een}{\end{enumerate}}
\newcommand{\bit}{\begin{itemize}}
\newcommand{\eit}{\end{itemize}}
\newcommand{\ol}{\overline}
\newcommand{\un}{\underline}
\newcommand{\lefq}{\lefteqn}
\newcommand{\ba}{\begin{array}}
\newcommand{\ea}{\end{array}}
\newcommand{\beqa}{\begin{eqnarray}}
\newcommand{\eeqa}{\end{eqnarray}}
\newcommand{\beqas}{\begin{eqnarray*}}
\newcommand{\eeqas}{\end{eqnarray*}}
\newcommand{\bfg}{\begin{figure}}
\newcommand{\efg}{\end{figure}}
\newcommand{\bds}{\begin{displaymath}}
\newcommand{\eds}{\end{displaymath}}
\newcommand{\btb}{\begin{tabbing}}
\newcommand{\etb}{\end{tabbing}}
\bc {
\textbf{   Position Dependent Mass Approach and Quantization for a Torus Lagrangian } } \ec

\vs{1cm}

\bc
{\it \"Ozlem Ye\c{s}ilta\c{s}{  \\
Department of Physics, Faculty of Science, Gazi University,
06500 Ankara, Turkey\\
\vspace{.16cm}

}} \ec \vs{1cm}

\begin{abstract}
We have shown that a Lagrangian for a torus surface can  yield second order nonlinear differential equations using the Euler-Lagrange formulation. It is seen that these second order nonlinear differential equations can be transformed into the nonlinear quadratic and Mathews-Lakshmanan equations using the position dependent mass approach developed by Mustafa \cite{om} for the classical systems. Then, we have applied the quantization procedure to the nonlinear quadratic and Mathews-Lakshmanan equations and found their exact solutions. 

\end{abstract}
\noindent {\bf keyword:}   classical Lagrangian, torus \\

\section{Introduction}
The bound states of a physical system have a fundamental importance for both quantum mechanics and classical theory. The curved surfaces and their bound states (such as curved strips \cite{surface}) have  attracted much attention and these models can be useful for the device modelling area. When we look at the surfaces with different geometries, a particle constrained to the surface of a torus has been studied in classical and quantum theories \cite{mott}, \cite{sree}, \cite{jh} ,\cite{bol}. At the same time, some of these models lead to applications in nano-ribbons  \cite{bol}. When it comes to the mathematical methods for the symmetries of the nonlinear systems, it is known that any second order non-linear differential equation admits eight parameter Lie point symmetries, thus they are solvable through point transformations \cite{ibr}, \cite{Lie1}, \cite{ben}, \cite{Lie2}, \cite{diff}. On the other hand, the  integrable and  super-integrable classical oscillators and their relationships with a nonlinear Li\'{e}nard type nonlinear equations are given in \cite{Car1}, \cite{Car2}, \cite{Car3}. The quantum version of a solvable  non-linear oscillator and the complete solutions can be found in terms of a family of orthogonal polynomials \cite{Car4}.

Because the nonlinear differential equations also appear in the solutions of torus Lagrangians and Hamiltonians, the main objective of this work is to search  nonlocal point transformations \cite{om} that can be available for our system. Thus, in this study, we  obtain a couple of non-linear differential equations for the torus surface using the Euler-Lagrange equations. The position dependent mass approach and the quantization procedure are applied to one of those  nonlinear differential equation. In fact, the authors have already found the exact solutions of the quantum  nonlinear quadratic   and Mathews-Lakshmanan oscillator models \cite{Car4} but we may follow a different approach for obtaining the solutions.

And this work is organized as follows; Section 2 contains the nonlinear equations obtained from the Lagrangian and the position dependent mass approach is adapted to the system. Moreover, adaptation of the toroidal equation to the examples of the nonlinear quadratic   and Mathews-Lakshmanan position dependent mass(PDM) nonlinear oscillator equations \cite{ML} are involved. Section 3 is introduced for the quantization of these oscillators and the exact solutions are also involved. Finally, the results are discussed in Section 4.
\section{Differential equations}
The metric  for the torus surface is given by
\begin{equation}\label{metric}
  ds^{2}=(c+a\cos v)^{2}du^{2}+a^{2}dv^{2}
\end{equation}
where $u, v$ are the coordinates. The Lagrangian can be written in terms of the metric tensor as
\begin{equation}\label{L}
  L(q,\dot{q})=g_{ij}\dot{q}_{i}\dot{q}_{j}
\end{equation}
where we take $q_1=u, q_2=v$. Then, the Lagrangian becomes
\begin{equation}\label{Lag}
  L=(c+a\cos v)^{2}\dot{u}^{2}+a^{2}\dot{v}^{2}.
\end{equation}
Using the Euler-Lagrange equations, we can derive the equations given below
\begin{eqnarray}\label{1}
  (c+a\cos v)\ddot{u}-2a\sin v~ \dot{u}\dot{v} &=& 0 \\
  a^{2}\ddot{v}+a(c+a\cos v)\sin v~ \dot{u}^{2} &=& 0.\label{2}
\end{eqnarray}
If we multipliy (\ref{1}) by $\dot{u}$ and integrate it, we get
\begin{equation}\label{3}
  \dot{u}=\frac{q_{1}}{(c+a\cos v)^{2}},
\end{equation}
  and
\begin{equation}\label{4}
  \dot{v}^{2}=\frac{q_{1}^{2}}{a^{2}(c+a\cos v)^{2}}+q_{2}.
\end{equation}
Here $q_{1}$ and $q_{2}$ are the integration constants. Differentiating each side of (\ref{3}) with respect to time and using (\ref{4}), we get
\begin{equation}\label{5}
  \ddot{v}+\frac{q_{1}^{2}}{a}\frac{\sin v}{(c+a\cos v)^{3}}=0.
\end{equation}
We may also note that for the value of $q_{2}=0$, we get
\begin{equation}\label{0}
  \ddot{u}=\frac{2q_{1}^{2}\sin v}{(c+a\cos v)^{4}}.
\end{equation}
We will continue with (\ref{5}) through this work.
\subsection{position dependent mass Lagrangian approach}
We will follow the notation given in \cite{om} here. Let us remind the Euler-Lagrange equation which is given by
\begin{equation}\label{06}
  \frac{d}{d\tau}\frac{\partial L}{\partial q}-\frac{\partial L}{\partial q}=0,
\end{equation}
and hence
\begin{equation}\label{6}
  \frac{d^{2}q}{d\tau^{2}}+\frac{\partial V}{\partial q}=0,
\end{equation}
where \cite{om}
\begin{equation}\label{60}
 \frac{dq(x)}{dx}=\sqrt{g(x)},~~ \frac{d \tau}{dt}=f(x),~~x=x(t).
\end{equation}
Under these transformations, (\ref{6}) becomes \cite{om}
\begin{equation}\label{7}
  \ddot{x}+\frac{1}{2}\frac{m^{'}(x)}{m(x)}\dot{x}^{2}+\frac{f(x)^{2}}{g(x)}\frac{\partial V}{\partial x}=0,
\end{equation}
where
\begin{equation}\label{70}
  \frac{m^{'}(x)}{m(x)}=\frac{g^{'}(x)}{g(x)}-2\frac{f^{'}(x)}{f(x)}.
\end{equation}
 We note that (\ref{7}) is also known as Li\'{e}nard type equation \cite{Lie1}. In \cite{om}, it is shown that the Euler-Lagrange equations (\ref{06}), (\ref{6}) and (\ref{7}) are invariant under the nonlocal point transformations:
\begin{equation}\label{8}
  L=L(q,\dot{q},\tau)=\frac{1}{2}(\frac{dq}{d \tau})^{2}-V(q)=\frac{1}{2}m(x)\dot{x}^{2}-V(x).
\end{equation}
Next, we will use the procedure explained above  to see how one can transform the system (\ref{5}) into (\ref{7}). Here, our task is to adapt (\ref{5}) to the nonlinear quadratic and Mathews-Lakshmanan oscillators.
\\
\emph{\textbf{Case 1}: \textbf{Nonlinear quadratic oscillator equation}}:\\
The nonlinear quadratic oscillator equation is written as \cite{Car1}, \cite{om}
\begin{equation}\label{pdm}
  \ddot{x}-\frac{2\lambda}{1+\lambda x}\dot{x}^{2}+\alpha^{2}x(1+\lambda x)=0.
\end{equation}
If (\ref{pdm}) and (\ref{7}) are compared, we obtain
\begin{equation}\label{9}
  m(x)=\frac{C_1}{(1+\lambda x)^{4}},
\end{equation}
\begin{equation}\label{10}
  V(x)=C_2-C_1\frac{\alpha^{2}(1+2x\lambda)}{2\lambda^{2}(1+x\lambda)^{2}}.
\end{equation}
Let us first consider  (\ref{6}) and (\ref{5}). Time derivatives can correspond to derivatives with respect to $\tau$ in (\ref{5}) and let $q=v$ be the nonlocal point transformation such that
\begin{equation}\label{11}
   \frac{d^{2}q}{d\tau^{2}}+\frac{\partial V}{\partial q}=0 \rightarrow  \frac{\partial V}{\partial q}=\frac{q_{1}^{2}}{a}\frac{\sin q}{(c+a \cos q)^{3}} \rightarrow V(q)=\frac{q_{1}^{2}}{2a^{2}(c+a\cos q)^{2}}+const.
\end{equation}
In order to find $q=q(x)$, we match $V(q)$ in (\ref{11}) and $V(x)$ in (\ref{10}), we obtain four roots of $q(x)$ as
\begin{equation}\label{12}
  q_{1,2}(x)=\pm \cos^{-1}(-\frac{c}{a}+\frac{i \lambda q_{1}}{\alpha a^{2}}\frac{1+\lambda x}{\sqrt{-C_1(1+2\lambda x)}}),
\end{equation}
\begin{equation}\label{q34}
  q_{3,4}(x)= \pm \cos^{-1}(\frac{-a^{2}c C_1 \alpha^{2}(1+2\lambda x)+q_{1}^{2}\lambda^{2}(1+\lambda x)^{2}
  \sqrt{-\frac{a^{2}C_1 \alpha^{2}(1+2\lambda x)}{q_{1}^{2}\lambda^{2}(1+\lambda x)^{2}}}}{a^{3}\alpha^{2}(C_1+2C_1 \lambda x)}).
\end{equation}
Using (\ref{70}) and (\ref{60}), it is obtained as \cite{om}
\begin{equation}\label{13}
  q(x)=\int \sqrt{m(x)} f(x) dx
\end{equation}
then, we get
\begin{equation}\label{14}
  f_{1,2}(x)=\pm \frac{q_{1}\lambda^{3}x(1+\lambda x)^{4}}{C_1 \alpha^{2}(1+2\lambda x)}(a^{2}(a^{2}-c^{2})C_1 \alpha^{2}(1+2\lambda x)-q_{1}^{2}\lambda^{2}F(x))^{-1/2}
\end{equation}
and
\begin{equation}\label{f34}
  f_{3,4}(x)=\mp \frac{a x \lambda^{2}\alpha^{2}(1+\lambda x)^{3}}{\sqrt{-a^{2}(a^{2}-c^{2})C_1\alpha^{2}(1+2x\lambda)-q_{1}^{2}\lambda^{2}(1+\lambda x)^{2}(F(x)+2)}}
\end{equation}
where $F(x)=-1+2c\sqrt{-\frac{a^{2}C_1 \alpha^{2}(1+2\lambda x)}{q_{1}^{2}\lambda^{2}(1+\lambda x)^{2}}}$. And the mass function in terms of the torus parameters can be given by
\begin{equation}\label{f35}
  m(x)=\frac{C_1}{a^{4}(1+2\lambda x)(1+\lambda x)^{-8}}\frac{(a^{2}(a^{2}-c^{2})C_1 \alpha^{2}(1+2\lambda x)-q_{1}^{2}\lambda^{2}F(x))^{-1/2}}
  {1-(-c/a+\frac{iq_{1}\lambda(1+\lambda x)}{a^{2}\alpha}\sqrt{-C_1(1+2\lambda x)})^{2}}.
\end{equation}
Thus, using (\ref{f35}) and (\ref{10}), (\ref{5}) can be written in the form of a Li\'{e}nard type equation in (\ref{7}).
\\
\emph{\textbf{Case 2}:\textbf{Mathews-Lakshmanan PDM nonlinear oscillator equation}}:\\
The Mathews-Lakshmanan oscillator equation is \cite{Car1}, \cite{om}
\begin{equation}\label{15}
  \ddot{x}-\frac{\lambda x}{1+\lambda x^{2}}\dot{x}^{2}+\frac{\omega^{2}x}{1+\lambda x^{2}}=0.
\end{equation}
In this case,  comparing (\ref{15}) and (\ref{7}), we have
\begin{equation}\label{16}
  m(x)=\frac{C_1}{1+\lambda x^{2}}
\end{equation}
and
\begin{equation}\label{17}
  V(x)=-\frac{C_1 \omega^{2}}{2\lambda(1+\lambda x^{2})}+C_2.
\end{equation}
And we obtain the functions $q(x)$ and $f(x)$ as
\begin{equation}\label{18}
  q(x)=\pm \cos^{-1}(-\frac{c}{a}\pm \sqrt{\frac{\lambda}{C_1}}\frac{q_{1}}{a^{2}\omega}\sqrt{1+\lambda x^{2}}),
\end{equation}
\begin{equation}\label{19}
  f(x)=\mp \frac{\lambda x (\frac{C_1}{1+\lambda x^{2}})^{1/2}}
  {C_1 a} \frac{1}{\sqrt{-\frac{1}{a^{2}}+\frac{C_1\omega^{2}(c^{2}-a^{2})}{q_{1}^{2}\lambda(1+\lambda x^{2})}\pm\frac{2c}{a^{2}}\sqrt{-\frac{a^{2}C_1\omega^{2}}{q_{1}^{2}\lambda(1+\lambda x^{2})}}}}.
\end{equation}
Finaly, the mass function can be written in terms of the torus parameters as
\begin{equation}\label{190}
  m(x)=\frac{q_{1}^{2}\lambda}{a^{2}\omega^{2}}\frac{-1/a^{2}+
  \frac{C_1\omega^{2}(c^{2}-a^{2})}{q_{1}^{2}\lambda(1+\lambda x^{2})}+
  \frac{2c\omega}{q_{1}}\sqrt{-\frac{C_1}{\lambda(1+\lambda x^{2})}}}{1-(-\frac{c}{a}+\frac{q_{1}}{a^{2}\omega}\sqrt{\frac{\lambda(1+\lambda x^{2})}{C_1}})^{2}}.
\end{equation}
\section{Quantization}
Now let us consider (\ref{4}) again.  If the Lagrangian is given for (\ref{4}) as
\begin{eqnarray}
  L(t,x,\dot{x}) &=& \frac{1}{2}(c+a \cos x)^{2}\dot{\theta}^{2}-\frac{k^{2}}{2a^{2}}-\frac{\ell}{2}(c+a \cos x)^{2}.
\end{eqnarray}
where we use $v \leftrightarrow x $. And the corresponding Hamiltonian  is
\begin{equation}\label{16}
  H=\frac{1}{2}\frac{p^{2}}{(c+a \cos x)^{2}}+\frac{q_{1}^{2}}{2a^{2}}+\frac{q_{2}}{2}(c+a \cos x)^{2}.
\end{equation}
The momentum operator can be also given by
\begin{equation}\label{17}
  \hat{p}=-i\frac{1}{c+a \cos x}\frac{\partial}{\partial x}.
\end{equation}
And the Hamilton operator becomes
\begin{equation}\label{18}
  H=-\frac{1}{2(c+a\cos x)^{2}}\frac{d^{2}}{dx^{2}}-\frac{a\sin x}{2(c+a\cos x)^{3}}\frac{d}{dx}+\frac{q_{1}^{2}}{2a^{2}}+\frac{q_{2}}{2}(c+a \cos x)^{2},
\end{equation}
where we take $\hbar=1$. To our knowledge, this Hamiltonian system is not an exactly solvable one. We will re-consider the models in \cite{Car1, Car2, Car3, Car4} which are quantum version of the classical oscillators discussed above. \\
\emph{\textbf{Case 1}: \textbf{Nonlinear quadratic oscillator equation}}:\\
The Lagrangian and the Hamiltonian relations for the corresponding sytem are given by \cite{Car1}
\begin{eqnarray}
  L &=& \frac{1}{2}\frac{C_1}{(1+\lambda x)^{4}}\dot{x}^{2}-C_2+C_1 \frac{\alpha^{2}(1+2\lambda x)}{2\lambda^{2}(1+\lambda x)^{2}} \\
  H &=& \frac{1}{2C_1}(1+\lambda x)^{4}p^{2} +C_2-C_1 \frac{\alpha^{2}(1+2\lambda x)}{2\lambda^{2}(1+\lambda x)^{2}}
\end{eqnarray}
and we may give the momentum operator for the quantization procedure as
\begin{equation}\label{19}
  p\rightarrow -i\sqrt{\frac{(1+\lambda x)^{4}}{C_1}}\frac{d}{dx}.
\end{equation}
The eigenvalue equation is
\begin{equation}\label{20}
  H\psi=E\psi,
\end{equation}
and (\ref{20}) brings us to
\begin{equation}\label{21}
  -\frac{1}{2C_1}(1+\lambda x)^{4}\psi^{''}(x)-\frac{\lambda}{C_1}(1+\lambda x)^{3}\psi^{'}(x)+\left(C_2-E-C_1 \frac{\alpha^{2}(1+2\lambda x)}{2\lambda^{2}(1+\lambda x)^{2}}\right)\psi(x)=0.
\end{equation}
Now we can give a point transformation as
\begin{equation}\label{22}
  z=\int \frac{dx^{'}}{(1+\lambda x^{'})^{2}}
\end{equation}
and we obtain
\begin{equation}\label{23}
  -\psi^{''}(z)+(2C_1C_2-2C_1E+C^{2}_1\alpha^{2}z^{2}+\frac{C^{2}_1\alpha^{2}}{\lambda}z)\psi(z)=0.
\end{equation}
The solution of the above equation is  well known and we apply a coordinate shift $z_{\varphi}=z+\varphi$ and we get
\begin{equation}\label{24}
  -\psi_{\varphi}^{''}(z_{\varphi})+(2C_1C_2-2C_1E+C^{2}_1\alpha^{2}z_{\varphi}^{2}-\frac{C^{2}_1\alpha^{2}}{4\lambda^{2}})\psi_{\varphi}(z_{\varphi})=0,
\end{equation}
where we take $\varphi=\frac{1}{2\lambda}$. Here we can write the solutions of (\ref{24}) as
\begin{equation}\label{25}
  E_{n}=\alpha(n+\frac{1}{2})+2C_2-\frac{C_1\alpha^{2}}{4\lambda^{2}},~~~~n=0,1,2,...
\end{equation}
and
\begin{equation}\label{26}
  \psi(z)=\frac{C_1 \alpha}{\sqrt{\pi}2^{n}n!}e^{-\frac{\xi^{2}}{2}}H_{n}(\xi),
\end{equation}
where $\xi=\sqrt{C_1\alpha}(z+1/2\lambda)$, $H_n(\xi)$ are the Hermite polynomials.\\
\emph{\textbf{Case 2}: \textbf{Mathews-Lakshmanan PDM nonlinear oscillator equation}}:\\
The Lagrangian and the Hamiltonian for the corresponding sytem are written as \cite{Car1}
\begin{equation}\label{27}
  L=\frac{1}{2}\frac{C_1}{1+\lambda x^{2}}\dot{x}^{2}+\frac{C_1\omega^{2}}{2\lambda (1+\lambda x^{2})}-C_2,
\end{equation}
\begin{equation}\label{28}
  H=\frac{1}{2}(1+\lambda x^{2})p^{2}-\frac{C_1\omega^{2}}{2\lambda (1+\lambda x^{2})}+C_2
\end{equation}
and the momentum operator for the quantization procedure is
\begin{equation}\label{29}
  p\rightarrow -i\sqrt{\frac{(1+\lambda x^{2})}{C_1}}\frac{d}{dx}.
\end{equation}
The form of the Hamiltonian operator becomes
\begin{equation}\label{30}
  H=-\frac{1}{2}\frac{(1+\lambda x^{2})}{C_1}\frac{d^{2}}{dx^{2}}-\frac{\lambda x}{2C_1}\frac{d}{dx}-\frac{C_1 \omega^{2}}{2\lambda(1+\lambda x^{2})}+C_2.
\end{equation}
Using the transformation given below
\begin{equation}\label{31}
  \psi(x)=(1+\lambda x^{2})^{\gamma}\phi(x)
\end{equation}
and the eigenvalue equation $H\psi=E\psi$, we have
\begin{equation}\label{32}
  -(1+\lambda x^{2})\phi^{''}(x)-(1+4\gamma)\lambda x \phi^{'}(x)+\frac{1}{ (1+\lambda x^{2})}\left(2C_1(C_2-E)\lambda-2\gamma\lambda^{2}-C^{2}_1\omega^{2}+(2C_1(C_2-E)\lambda^{2}-4\gamma^{2}\lambda^{3})x^{2}\right)\phi(x)=0.
\end{equation}
Changing variable $z=-i\sqrt{\lambda }x$ in above equation leads to
\begin{equation}\label{33}
  (1-z^{2})\phi^{''}(z)-(1+4\gamma)z\phi^{'}(z)+\frac{1}{1-z^{2}}(\frac{2C_1(C_2-E)}{\lambda}-2\gamma-\frac{C^{2}_{1}\omega^{2}}{\lambda^{2}}+
  (4\gamma^{2}-\frac{2C_1(C_2-E))}{\lambda})z^{2})\phi(z)=0
\end{equation}
and using $(z^{2}-1)^{1/4-\gamma}F[z]$ in (\ref{33}), we obtain
\begin{equation}\label{34}
   (1-z^{2})F^{''}(z)-2zF^{'}(z)+
   \frac{1}{1-z^{2}}(\frac{2C_1(C_2-E)}{\lambda}-2\gamma-
   \frac{C^{2}_{1}\omega^{2}}{\lambda^{2}}+\frac{4\gamma-1}{2}+(4\gamma^{2}-\frac{2C_1(C_2-E)}{\lambda}+\frac{1-16\gamma^{2}}{4})z^{2})G(z)=0.
\end{equation}
One can transform (\ref{34}) into a Legendre differential equation, if  $\omega$ is constrained as
\begin{equation}\label{35}
 \omega=\pm i\frac{\lambda}{2C_1},
\end{equation}
thus, we obtain
\begin{equation}\label{36}
  (1-z^{2})F^{''}(z)-2zF^{'}(z)+\nu(\nu+1)F(z)=0,
\end{equation}
where
\begin{equation}\label{37}
  \nu(\nu+1)=-\frac{1}{4}+\frac{2C_1(C_2-E)}{\lambda}.
\end{equation}
Thus, the eigenvalues and the solutions are given respectively as
\begin{equation}\label{38}
  E_{\nu}=C_2 -\frac{\lambda}{8C_1}(2\nu+1)^{2},~~~~\nu=0,1,2,...
\end{equation}
 \begin{equation}\label{39}
   \phi_{\nu}(z)=(z^{2}-1)^{1/4-\gamma}P_{\nu}(z)
 \end{equation}
where $P_{\nu}(z)$ are the Legendre polynomials. And the function $\psi(x)$ becomes
\begin{equation}\label{40}
  \psi_{\nu}(x)=N(1+\lambda x^{2})^{1/4}P_{\nu}(-i\sqrt{\lambda }x).
\end{equation}



\section{Conclusions}
We have shown that a classical Lagrangian system for a torus surface can be transformed into a Li\'{e}nard equation using position dependent mass approach. In this manner, we have obtained the necessary coordinate and mass functions appearing in Euler-Lagrange equations to transform a toroidal equation to the well-known systems which are nonlinear quadratic oscillator  and Mathews-Lakshmanan  nonlinear oscillator equations. Finally we have discussed the exact solutions of the quantized eigenvalue equations of the corresponding classical oscillators. The quantum nonlinear quadratic oscillator Hamiltonian solutions are obtained in terms of Hermite polynomials with harmonic oscillator energy eigenvalues.  Using necessary transformations we have solved quantum Mathews-Lakshmanan  nonlinear oscillator whose solutions are given in terms of Legendre polynomials with an imaginary argument. In case of $C_2 < \frac{\lambda}{8C_1}(2\nu+1)^{2}$, the spectrum is negative which can be a difference between our results and the work in \cite{ML}. Here, another interesting result is an imaginary frequency in (\ref{35}) if $\lambda$ and $C_1$ are both real numbers. On the other hand, if one the constants $\lambda$ or $C_1$ is chosen as an imaginary parameter in order to make the frequency  as a real number, this choice  leads to a complex energy spectrum. Finally, we note that the transformation of (\ref{18}) to the eigenvalue equations (\ref{21}) and (\ref{32}) in examples 1 and 2 cannot be shown in the recent paper and this procedure will be in progress.

\end{document}